\documentclass[12pt]{iopart}

\usepackage[
  left=30mm, 
  right=30mm,
  includehead,
  headheight = 30mm,
  footskip = \dimexpr\headsep+\ht\strutbox\relax,
  tmargin = 0mm,
  bmargin = \dimexpr25mm+2\ht\strutbox\relax,
]{geometry}

\usepackage{graphicx}% Include figure files
\usepackage{subfig} % subfigures
\usepackage{epstopdf} % should this be removed by the journal??

\usepackage{iopams}  
\usepackage{amssymb}

\usepackage{threeparttable,array,booktabs,calc}

\usepackage{hyperref}

\usepackage[sort&compress, numbers]{natbib} % square brackets
\AtBeginDocument{}

\begin{document}

\title[]{Zeeman deceleration of electron-impact-excited metastable helium atoms}
\author{Katrin Dulitz$^1$, Atreju Tauschinsky$^1$ and Timothy P Softley$^1$\footnote{Corresponding author: tim.softley@chem.ox.ac.uk}}
\address{$^1$Department of Chemistry, University of Oxford, Chemistry Research Laboratory, 12 Mansfield Road, Oxford, OX1 3TA, United Kingdom}

\begin{abstract}
We present experimental results that demonstrate -- for the first time -- the Zeeman deceleration of helium atoms in the metastable 2$^3$S$_1$ state. A more than 40~$\%$ decrease of the kinetic energy  of the beam is achieved for deceleration from 490~m/s to a final velocity of 370~m/s. Metastable atom generation is achieved with an electron-impact-excitation source whose performance is enhanced through an additional discharge-type process which we characterize in detail. 
Comparison of deceleration data at different electron beam pulse durations  confirms that a matching between the initial particle distribution and the phase-space acceptance of the decelerator is crucial for the production of a decelerated packet with a well-defined velocity distribution. The experimental findings are in good agreement with three-dimensional numerical particle trajectory simulations.
\end{abstract}

\vspace{2pc}

\section{\label{sec:intrometastables}Introduction}

Atoms and molecules in excited states are referred to as being `metastable', when their tendency to return to the electronic ground state is strongly reduced, often because electric dipole transitions to lower-lying states are formally forbidden. This results in natural lifetimes longer than 100~$\mu$s and thus allows for the study of metastables in time-of-flight experiments \cite{Trajmar1992, Gay1996, Hotop1996}. 
Atoms and molecules in metastable states play an important role in the Earth's upper atmosphere, where they are formed through collisions with photoelectrons and participate in various (de-)excitation and ionization cycles \cite{Rundel1974, Frederick1977, Torr1979}. In this part of the atmosphere, metastable species are observed through their radiative emission, e.g., atomic nitrogen in the metastable $^2$D state gives rise to an emission at 520~nm, which forms part of the visible aurora in the high-latitude regions of the Arctic and Antarctic \cite{Rees1985}. Radiative emission from metastable states is an important resource in astrophysics, used to determine the temperature, density and chemical constituents of planetary atmospheres \cite{Kim2001, Galand2002}. Chemical reactions involving metastable species are often barrierless and exergonic, and are thus prevalent in diverse temperature and density regimes,  such as in the Earth's upper atmosphere \cite{Zipf1969, Takayanagi1970, Chakrabarti1998, Wayne2000}, in the atmospheres of planets and their satellites \cite{Wayne2000} and in combustion and plasma processes \cite{Baulch2005, Herron2001} where they are efficiently produced, e.g., by electron impact or charge transfer.

Metastable species often possess one or more unpaired electron spins so that they become suitable systems of choice for Zeeman deceleration, an experimental technique relying on the fast switching of inhomogeneous magnetic fields to remove kinetic energy from a supersonic beam of paramagnetic species. 
Zeeman deceleration has already been successfully demonstrated for supersonic beams of Ne(3$^3$P$_{2}$) \cite{Narevicius2008, Wiederkehr2011}, Ar(4$^3$P$_{2}$) \cite{Trimeche2011} and He$_2$($a^{3}\Sigma_{u}^{+}$) \cite{Motsch2014}.
Metastable rare gases, including He(2$^3$S$_1$), have also been cooled down into the $\mu$K regime \cite{Vassen2012} using laser cooling, and the magnetic moment is an important property for magnetic trapping.  The development of sources of cold metastable atoms of sufficient density has potential applications in cold chemistry, given the barrierless nature of reactions involving these species.

In this paper we report on the Zeeman deceleration of He(2$^3$S$_1$) for the first time. Even though we can only reach mK temperatures with a Zeeman decelerator, the relatively high magnetic-moment-to-mass ratio  of He(2$^3$S$_1$), and the good signal intensities make it a favorable test ground for Zeeman deceleration experiments with an electron-impact source.

Owing to the very low atomic mass, the velocity of a  supersonic beam of pure helium gas from a room-temperature source is on the order of 1800~m/s (see below) thus requiring dilution in a heavier carrier gas or cryogenic cooling to reduce the initial velocity of the beam so that the atomic system can be efficiently addressed with a Zeeman decelerator. Previous work with a discharge source \cite{Raunhardt2009} showed that atomic helium in the metastable 2$^3$S$_1$ state was entirely quenched upon seeding in a heavy carrier gas, possibly due to Penning ionization processes. An alternative approach developed by Motsch et al for metastable He$_2$ is to use a cryogenically cooled discharge source,  whose design is quite involved \cite{Motsch2014}. 

In our work we demonstrate and characterize the use of an electron-impact source for metastable atom generation. We show that the pulse duration for electron-impact excitation needs to be matched to the phase-space acceptance of the decelerator in order to attain a good contrast between the decelerated and undecelerated parts of the beam. In the second part of the paper,  we show that the production of He(2$^3$S$_1$) can be significantly increased if the source is operated in a regime where both discharge and electron-impact excitation can be achieved. This work lays the foundation for further metastable deceleration experiments and, eventually, the study of cold chemical reactions using Zeeman-decelerated metastable atoms and molecules.

\section{\label{sec:setup}Experimental}

\begin{figure}[ht!]
\centering
\includegraphics[width=0.95\textwidth]{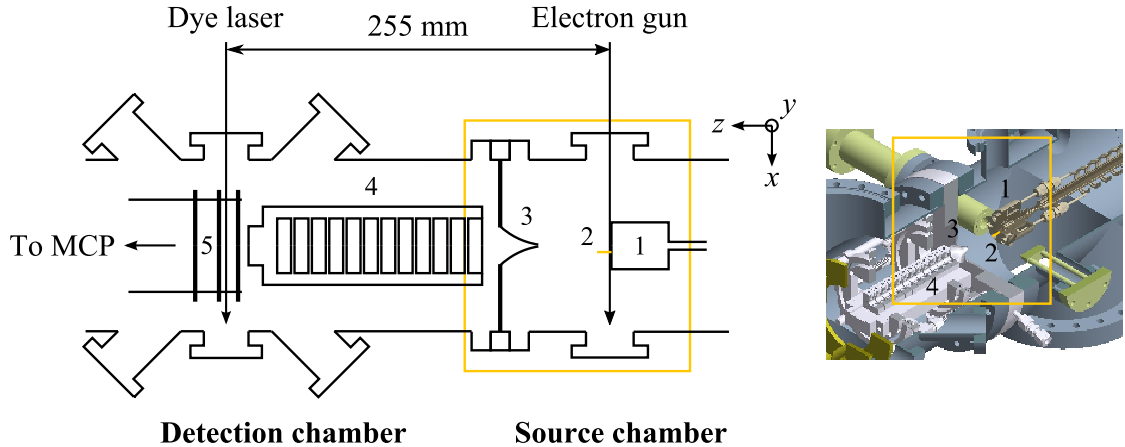}
\caption{Sketch of the experimental setup, not to scale. 1: Pulsed valve, 2: Pin, 3: Skimmer, 4: Zeeman decelerator, 5: Extraction plates.}
\label{fig:sketchchamber}
\end{figure}

The experimental arrangement is similar to our previous work on deceleration of ground-state hydrogen atoms \cite{Dulitz2014}, with the main difference being the use of an electron-impact source for the production of metastable atoms and molecules. In Figure~\ref{fig:sketchchamber}, a schematic illustration of the experimental setup is shown along with a CAD drawing of the source region. The molecular beam apparatus consists of a chamber for the generation of supersonic beams, and a chamber containing both the Zeeman decelerator and a Wiley-McLaren-type ion-time-of-flight detection system. The chambers are separated by a skimmer (Beam Dynamics, 2.0~mm orifice diameter), and differential pumping is achieved using two turbo pumps. During experiments, the pressures are typically at 1$\times$10$^{-5}$~mbar in the source chamber and at 4$\times$10$^{-8}$~mbar in the detection chamber. Without a gas load, pressures of 2--5$\times$10$^{-9}$~mbar are obtained in both chambers. 

The electron-impact excitation source used here provides a well-defined electron energy distribution and low energy spread compared to a discharge-type setup. The electron source, described  below, is mounted perpendicular to the pulsed valve and aimed close to the valve orifice to increase metastable production. Cross sections for metastable excitation are highest just above threshold ($\leq$~10$^{-17}$~cm$^2$) and level off towards higher electron energies \cite{Fabrikant1988}. However, we found optimum working conditions for metastable generation at kinetic energies of 100~eV (the maximum attainable kinetic energy for this source) due to considerably higher electron emission currents under these conditions. For deceleration experiments, the electron gun is pulsed for a duration of either 20~$\mu$s or 50~$\mu$s. A 10~mm long stainless steel pin with a 1~mm diameter is mounted on the valve plate along the axis of the electron beam (see Figure \ref{fig:sketchchamber}). Positive biasing of the pin (typically $U_{\mathrm{bias}}$~=~150~V) leads to an avalanche of secondary electrons in Ar, Kr and N$_2$ gas which have sufficient kinetic energy to enhance metastable formation (see Section \ref{sec:discharge}).

\subsection{\label{sec:egun}Electron gun}

The electron-impact source used for these experiments 
is based on a standard, three-lens electron-optics system, 
similar to the design described by Erdman and Zipf \cite{Erdman1982}. An yttria-coated iridium filament (ES-526, Kimball Physics, 3.9~A) is used as a high-intensity (emission current of 3--5 mA), thermionic electron emitter which is suitable for relatively low vacuum conditions, as occur close to the pulsed valve. An energy spread of approximately 0.6~eV is inferred from a near-threshold electron-impact excitation measurement of atomic helium. The use of magnetic shielding, with a mu-metal tube around the electrostatic lenses, and the reduction of stray magnetic fields, e.g, from cold-cathode pressure gauges, were crucial for the realization of an intense electron beam.

The electron gun is operated in pulsed mode to attain a well-defined phase-space window that is advantageous for Zeeman deceleration experiments (see Section \ref{sec:metastableHe}). Pulsed operation also provides a significantly larger electron current than in continuous mode, which is probably due to the influence of patch fields on continuous electron emission \cite{Murray1999}.
The beam current is detected on a positively biased metal pin that is mounted on the valve body (Figure~\ref{fig:sketchchamber}), and  the signal is then amplified and monitored with an oscilloscope. The current detector was not calibrated; since the collection efficiency is $<$~1, the measured values provide a lower estimate of the actual electron beam current. 

\begin{figure}[!ht]
\centering
\includegraphics[width=0.47\textwidth]{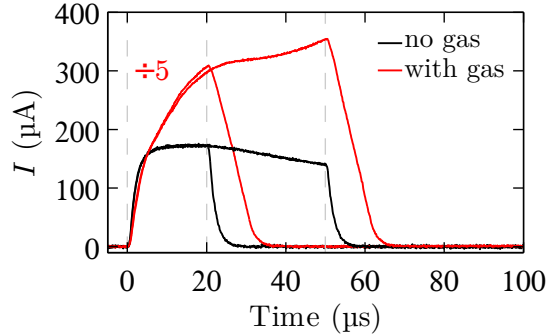}
\caption{Electron current, $I$, detected at the pin as a function of time for a 1:3 He/Ar mixture ($p_0$~=~6~bar, $T_0$~=~301~K, $U_{\mathrm{bias}}$~=~150~V) with or without gas present, as specified in the legend. Traces were recorded for electron gun pulse durations $\Delta t_\mathrm{e}$~=~20~$\mu$s and 50~$\mu$s, which are significantly shorter than the supersonic gas beam pulses. Triggers to the electron gun pulser are indicated with gray dashed lines. Signals recorded with the gas pulse present are scaled down by a factor of 5 for visibility.}
\label{fig:egunchar}
\end{figure}

Figure \ref{fig:egunchar} shows typical temporal profiles of the electron current for pulse durations, $\Delta t_\mathrm{e}$, of 20~$\mu$s and 50~$\mu$s. When the pulsed valve (for the gas beam) is not operated, the detected electron current shows a nearly rectangular shape for both pulse durations, with 5~$\mu$s rise and fall times. In the presence of the supersonic gas beam pulse, the measured intensity profile can be significantly different from that, as highlighted for a 1:3 He/Ar mixture in Figure \ref{fig:egunchar}. We attribute this difference to the detection of secondary electrons produced during the electron-impact excitation of the gas (see Section \ref{sec:discharge}). Longer pulse durations result in an almost linear increase in metastable signal (not shown) which is in accordance with the higher number of emitted electrons.

\subsection{\label{sec:setupchar}Beam velocity and intensity measurements by direct MCP detection}
\begin{figure}[!ht]
\centering
\subfloat[]{\label{MCP0V}%
\includegraphics[height=0.219\textheight]{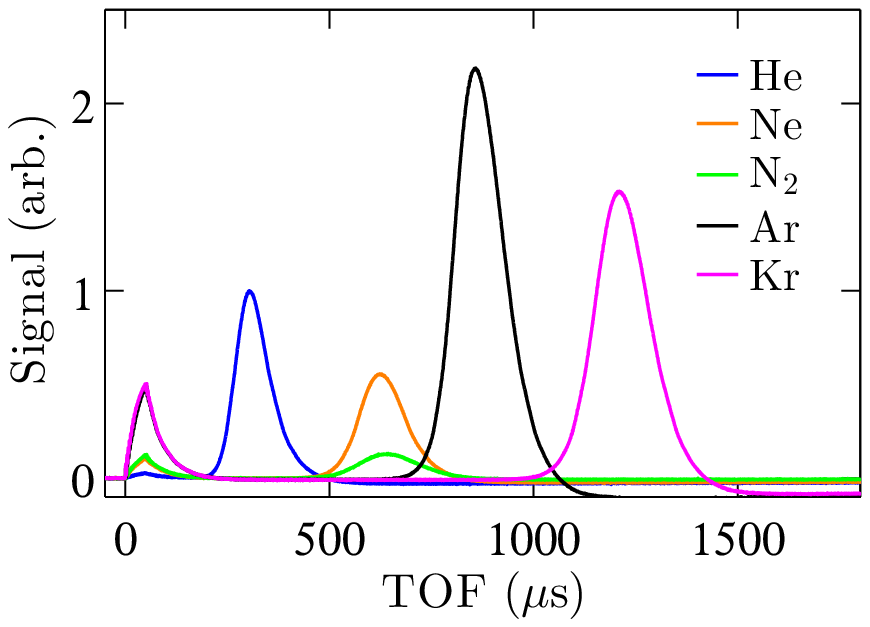}}
\hspace{10pt}
\subfloat[]{\label{varyp0V}%
\includegraphics[height=0.219\textheight]{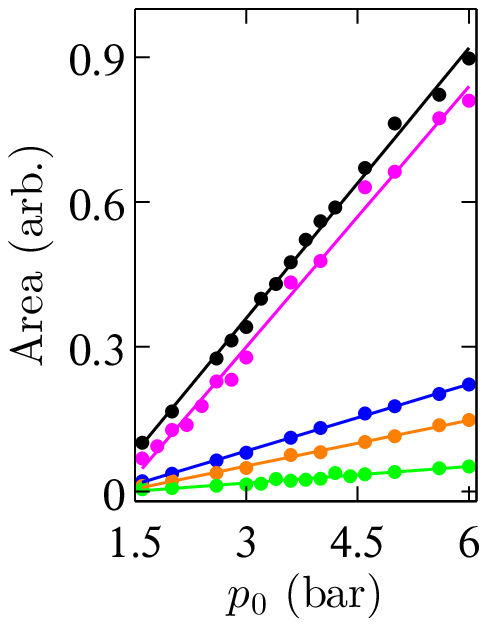}}
\caption{(a) TOF profiles of metastable rare gas atoms and molecular nitrogen (see legend, $T_0$~=~302~K and $p_0$~=~6.0~bar) from Auger electron detection on the MCPs (-1650~V). The peaks at earliest arrival time are from light emission during electron-impact excitation. Negative signal intensities are due to an artifact of the detection method (charge amplifier). The data are normalized to the metastable He signal intensity. (b) Change in integrated TOF signal as a function of source pressure; legend as in (a). Each data set is fitted with a linear regression. The pin is grounded during the measurements.}
\label{fig:MCPmetastables}
\end{figure}

The detection of metastable atoms and molecules by Auger electron ejection from a metal-containing surface, such as an MCP detector, is possible if the particles' kinetic energy is above the work function $\Phi$ of the metal \cite{Hotop1996}. For an MCP detector, $\Phi$ is approximately 5.4~eV (lead glass) \cite{Melamid1972}
, so that, as illustrated by time-of-flight (TOF) profiles in Figure \ref{MCP0V}, metastable rare gases and N$_2$(A$^3\Sigma_{u}^{+}$) induce such an emission signal. For these measurements, the repeller and extractor plates are set at constant positive voltages of 750~V and 500~V, respectively, to reject ions from impinging on the MCP surface. The total flight path between the excitation region and the detector is the same as in the deceleration measurements, i.e., the particles always pass through the Zeeman decelerator (not operated here) and the TOF setup before detection (overall distance $\approx$~500~mm). As can be seen from Table \ref{tab:supersonicvel}, the measured beam velocities of the various species are in good agreement with theoretical estimates. 

\begin{table}[hbt]
\centering
\caption{Experimental flow velocities for metastable atoms and molecules from time-of-flight measurements and comparison with theoretical estimates \cite{Christen2008}.  Beam velocities were extracted from the TOF profiles, assuming a Maxwell Boltzmann distribution and taking into account the finite pulse duration for electron-impact excitation ($\Delta t_\mathrm{e}$~=~50~$\mu$s) and assuming a well-collimated supersonic beam \cite{Haberland1985}.}\label{tab:supersonicvel}
\begin{threeparttable}
\begin{tabular}{ccc}
\toprule[1.7pt]
Primary Metastable & Theory & Experiment \\
Species \cite{Gay1996} & $v_{\infty}$ (m/s) & $v_{\mathrm{exp}}$ (m/s) \\
\midrule[1.7pt]
He(2$^1$S$_0$), He(2$^3$S$_1$) & 1772 & 1800\\
Ne(3$^3$P$_{0,2}$) & 789 & 840\\
N$_2$(A$^3\Sigma_{u}^{+}$) & 792 & 840\\
Ar(4$^3$P$_{0,2}$) & 561 & 610\\
Kr(5$^3$P$_{0,2}$) & 387 & 430\\
\bottomrule[1.7pt]
\end{tabular}
\end{threeparttable}
\end{table}

Integration of the time-of-flight peaks gives the relative intensity of metastables transmitted to the detector. The integrated signal follows a linear dependence with respect to source pressure (Figure \ref{varyp0V}) for all the gases, which can be rationalized through the larger number of atoms/molecules available for excitation towards higher pressures. However, the fitted lines have different slopes indicating that mechanisms such as  charge transfer and energy transfer may also be important for the excitation processes.

\subsection{\label{sec:setupdecel}REMPI detection of metastable atoms}

The MCP detection method described above does not allow for a discrimination between different metastable states. In addition to that, the MCP signal obtained when seeding in a heavy carrier gas like Ar or Kr is dominated by the contribution of the latter species. Therefore, the Zeeman-decelerated He atoms in the 1$s$2$s$ $^3$S$_1$ state are detected using a (1+1) resonance enhanced multi-photon ionization (REMPI) process using the strong transition at 388.97~nm via the 1$s$3$p$ $^3$P$_{0,1,2}$ state \cite{Dunning1974}. Approximately 10~mJ/pulse (10~Hz, 5~ns) of laser light  at this wavelength is produced by a 355 nm pulsed dye laser system (Quanta-Ray PDL-3, Spectra-Physics, Exalite 389 dye). 
A lens ($f$~=~150~mm) is used to focus the detection laser beam into the vacuum chamber. Again, a continuous voltage is applied to the extraction plates to prevent the detection of ions produced by the electron-impact excitation process on the MCP detector.

\section{\label{sec:metastableHe}Zeeman deceleration of metastable helium atoms}
A supersonic beam of He atoms was generated by seeding He in Ar carrier gas (He/Ar mixing ratio of 1:3, $p_0$~=~4.5~bar) and cooling the valve body to a temperature of 143~K.  The mean initial velocity of 520~m/s, and the longitudinal and transverse temperatures of $T_z$~=~0.7~K and $T_r$~=~0.1~K, are deduced from the position and shape of the TOF signal in comparison with trajectory simulations. By choosing a deceleration sequence we can select atoms either from the center of this distribution or from the wings to be decelerated, allowing us to tune the initial velocity between 490~m/s and 520~m/s.

For the simulations, the electron beam is assumed to have a 2~mm diameter and the detection laser beam is chosen to have a diameter of 2.5~mm. Several authors have calculated and experimentally demonstrated the deflection of metastable helium atoms in a transverse geometry for electron-impact \cite{Brutschy1977, Tommasi1992}. To account for such a deflection by the electron beam, the metastable He atoms are simulated with an additional off-axis velocity of 30~m/s. This value was not confirmed experimentally, but it tended to provide a better agreement with the experimental results. 

The experimental parameters used for the Zeeman decelerator are similar to the previous studies on ground-state H atoms \cite{Dulitz2014}, with all 12 deceleration coils  operated at a current of 243~A, resulting in rise and fall times of about 8~$\mu$s.

\subsection{\label{sec:psmatching}Phase-space matching}

Compared to the Zeeman deceleration experiments with H atoms, the influence of the finite pulse duration for electron impact was one of the major questions to be addressed for metastable He. While excimer laser photolysis of NH$_3$ typically generates H atoms in a time interval of about 10~ns, the pulse duration for an electron-impact source can be tuned over a wide range (Section \ref{sec:egun}), which offers the chance to completely fill the 6~D phase-space volume accepted by the Zeeman decelerator. However, it also introduces uncertainties arising from less well-defined source conditions, e.g., from the non-uniform electron pulse shape and the excitation in different parts of the initial gas pulse. In addition to that, to obtain a decelerated beam with useful properties, the pulse duration for electron-impact excitation needs to be matched to the phase-space acceptance of the Zeeman decelerator to prevent the transmission of particles outside the accepted phase-space volume.

The effect of phase-space matching, i.e., the adjustment of the initial phase-space distribution of the beam to the phase-space acceptance of the decelerator, has already been investigated in the context of Stark deceleration and electrostatic trapping \cite{Bethlem2002, Hudson2004, van2005b}. However, there has not been a systematic study on the width of the incoming particle distribution thus far. In discharge experiments with OH radicals, the use of longer pulse durations  caused a broadening of the longitudinal velocity distribution and an increase in the measured rotational temperature of the OH beam, so that very short excitation widths (2~$\mu$s) were adopted in general \cite{Hudson2006b}. 

\begin{figure}[ht!]
\centering
\includegraphics{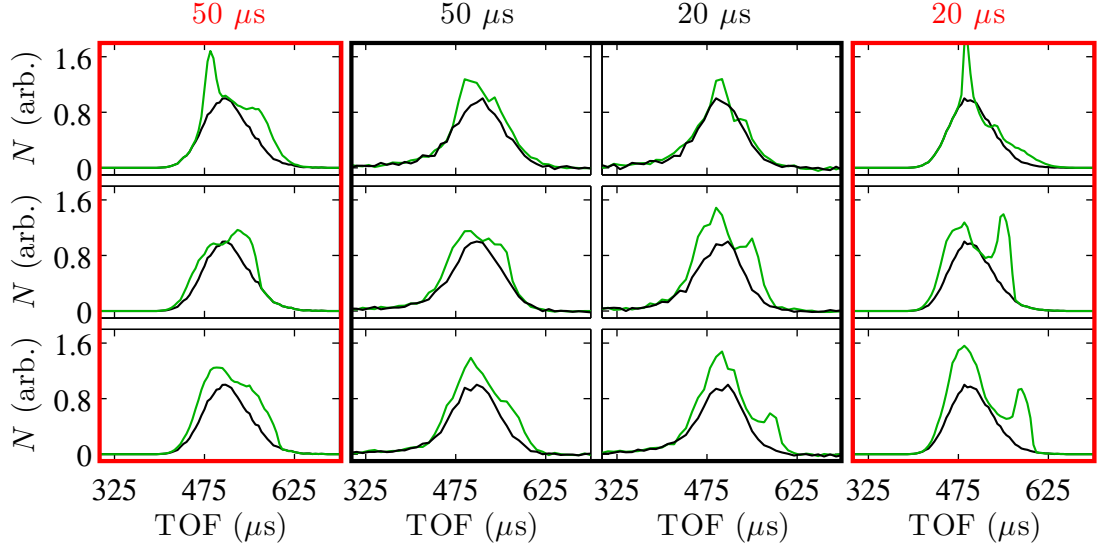}
\caption{Experimental (black frame) and simulated (red frames) TOF traces for Zeeman deceleration/acceleration of He(2$^3$S$_1$) from 505~m/s to 590~m/s ($\bar{a}_{z}/\bar{a}_{z,m}$~=~1.0, top panel), to 450~m/s ($\bar{a}_{z}/\bar{a}_{z,m}$~=~-0.5, middle panel) and to 390~m/s ($\bar{a}_{z}/\bar{a}_{z,m}$~=~-1.0, bottom panel). Traces were taken on a two-shot basis (black traces: Zeeman decelerator off, green traces: Zeeman decelerator on) at pulse durations of 20~$\mu$s and 50~$\mu$s for the electron gun, as indicated above the figure. The black traces were used for normalization. Here, $p_0$~$\approx$~4.1~bar, $T_0$~=~143~K.}
\label{fig:TOFHe}
\end{figure}

To characterize our system, Zeeman deceleration measurements were carried out at two different pulse durations for electron-impact excitation, 20~$\mu$s and 50~$\mu$s. Figure \ref{fig:TOFHe} contrasts the measured time-of-flight profiles (TOF) for experiments in which the pulse sequence was set to accelerate/decelerate the synchronous particle from an initial He(2$^3$S$_1$) velocity of 505~m/s to 590~m/s (top panel), to 450~m/s (middle panel) and to 390~m/s (bottom panel).  The profiles are compared with the output from 3~D particle trajectory simulations \cite{Dulitz2014}. In this case, the time-of-flight is defined as the delay between the rising edge of the electron pulse and the detection laser beam.

Figure \ref{fig:TOFHe} illustrates that the peak corresponding to the decelerated packet is shifted to later arrival times, hence lower particle velocity, as the mean longitudinal acceleration per coil, $\bar{a}_{z}/\bar{a}_{z,m}$ (see \cite{Dulitz2014a} for a detailed discussion of this parameter), is set to more negative values. 
Compared to a pulse duration of 50~$\mu$s, the contrast between the decelerated and undecelerated parts of the beam is much more pronounced at the shorter, 20~$\mu$s pulse duration. This difference can be explained through a comparison with the longitudinal phase-space acceptance for Zeeman deceleration \cite{Dulitz2014a}. 

\begin{figure}[ht!]
\centering
\includegraphics{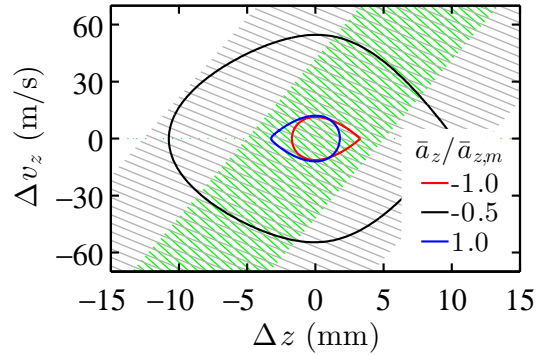}
\caption{Separatrices in longitudinal phase space for the Zeeman deceleration of He(2$^3$S$_{1}$) at different mean longitudinal accelerations, $\bar{a}_{z}/\bar{a}_{z,m}$, as indicated in the legend. Green and gray hashes indicate the extent of the initial particle distributions expected for electron gun pulse durations of 20~$\mu$s and 50~$\mu$s after a flight time of 120~$\mu$s (approximate switch-on time for the first deceleration coil), respectively.}
\label{fig:fishHe}
\end{figure}

Figure \ref{fig:fishHe} shows longitudinal separatrices, determined as outlined in \cite{Dulitz2014a}, which would be expected for the chosen deceleration pulse sequences and compares them with the extent of the initial particle distributions, corresponding to electron gun pulse durations of 20~$\mu$s (green hashes) and 50~$\mu$s (gray hashes). The latter are shown at the approximate switch-on time for the first deceleration coil, so that the distributions are tilted owing to free flight from the source region to the first deceleration coil. From the beam temperature, the longitudinal velocity distribution has a full-width-at-half maximum of about 87~m/s; the spatial spread should be approximately uniform (not shown in the figure). Figure \ref{fig:fishHe} illustrates that the phase-space extent from a 50~$\mu$s electron pulse is much wider than the actual longitudinal acceptance of the Zeeman decelerator, especially at $|\bar{a}_{z}|/\bar{a}_{z,m}$~=~1.0. While a long pulse duration ensures that virtually all particles in the phase-space region are addressed by the pulsed magnetic fields, it also leads to the partial deceleration/acceleration of particles outside the accepted phase-space volume. As our Zeeman decelerator is relatively short, these particles can reach the detection region, causing an overall blurring of the decelerated signal (as seen in the TOF profiles in Figure \ref{fig:TOFHe}), and a much broader velocity distribution than expected from phase-space calculations. These results show that, even though much higher metastable signal intensities can be achieved through longer pulse durations, the actual number of decelerated particles inside the phase-space volume is not increased any further. 

Both the peak positions and the peak shape of the TOF profiles obtained from trajectory simulations are in good agreement with the experimental results, and reflect the observed blurring of the signal at longer pulse durations. The effect of phase-space matching also becomes obvious when the incoupling time, i.e., the relative timing between electron-impact excitation and the beginning of the Zeeman deceleration sequence, is scanned (not shown). Experiments at even shorter pulse durations were not successful owing to very low signal intensities.

\subsection{\label{sec:limitations}Experimental limitations}

\begin{figure}[ht!]
\centering
\includegraphics{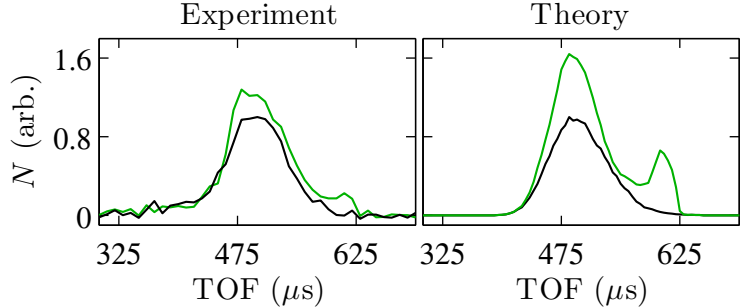}
\caption{Experimental and simulated TOF traces for Zeeman deceleration of He(2$^3$S$_1$) from 490~m/s to 370~m/s ($\bar{a}_{z}/\bar{a}_{z,m}$~=~-1.0). Color code as in Figure \ref{fig:TOFHe}. Traces were taken at a pulse duration of 20~$\mu$s for the electron gun. Here, $p_0$~=~5.4~bar, $T_0$~=~146~K.}
\label{fig:lowestvzHe}
\end{figure}

Figure \ref{fig:lowestvzHe} shows an experimental and simulated TOF trace for the lowest velocity attained with the current setup. In this case, an initial velocity of 490~m/s was chosen for deceleration and a final velocity of 370~m/s was achieved, corresponding to a more than 40~$\%$ decrease of the particles' kinetic energy. The signal for the decelerated peak (at 610~$\mu$s, green trace) is clearly separated from the arrival times of the undecelerated beam (black trace). However, the attained signal intensity is rather low, because the selected velocity for deceleration is not in the center of the initial velocity distribution (520~m/s). Again, the agreement with trajectory simulations is reasonable.
%In general, the level of agreement with trajectory simulations is only qualitative. 
Deviations between experimental and simulated data may be related to the finite duration
%This is mainly due to the ill-defined characteristics of the initial supersonic beam, e.g., owing to finite duration 
and the non-uniform shape of the electron-impact excitation pulse. Moreover, there are large uncertainties arising from the discharge-like excitation (see below) and the transverse geometry of the electron-gun setup. In addition to that, the signal-to-noise ratio for this data is limited, mainly due to the rather low metastable flux.

\section{\label{sec:charmetastables}Further characterization of the metastable source and implications for metastable helium production}

This Section gives a more detailed characterization of the electron-beam source. Evidence will be provided that metastable generation can be strongly increased through an additional discharge-type excitation process for certain gases. Through mixing with an appropriate carrier gas, this mechanism can also be used to increase the generation of He(2$^3$S$_{1}$).

\subsection{\label{sec:discharge}Transition from electron-impact to discharge-type excitation}

\begin{figure}[!ht]
\centering
\subfloat[]{\label{varyp}%
\includegraphics[height=0.205\textheight]{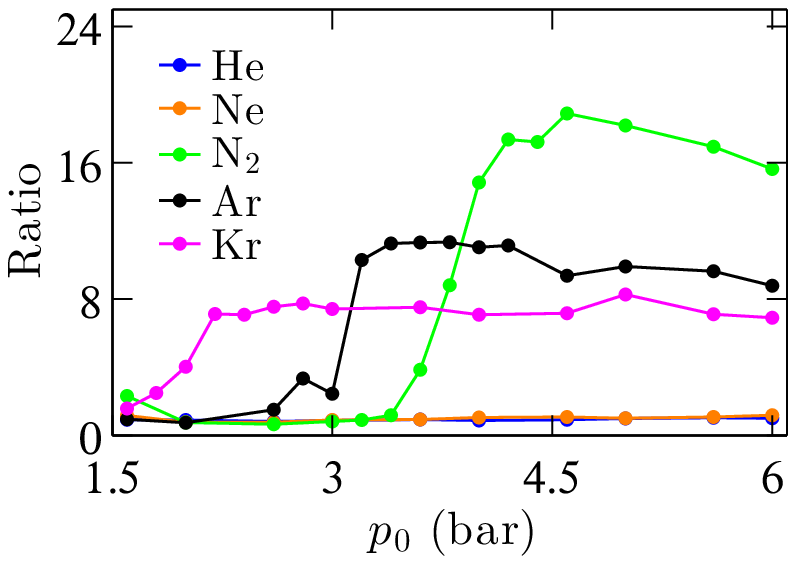}}
\hspace{10pt}
\subfloat[]{\label{varyUbias}%
\includegraphics[height=0.21\textheight]{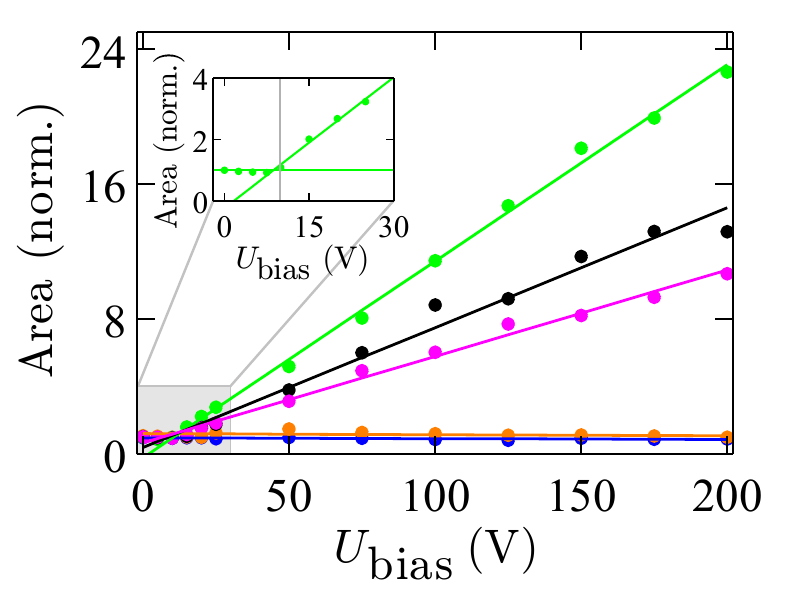}}
\caption{(a) Ratio of the integrated MCP signal intensity with and without a 150~V bias voltage as a function of source pressure. (b) Change of integrated MCP signal intensity as a function of bias voltage for different gases (same legend as in (a), $T_0$~=~301~K and $p_0$~=~5.6~bar) and linear fits. The onset of the signal increase for N$_2$ is shown in the inset; the N--N bond-dissociation energy (in eV) is marked with a gray vertical line. Signals are normalized to the results obtained at $U_{\mathrm{bias}}$~=~0~V. }
\label{fig:varybias}
\end{figure}

Measurements using the valve at room temperature, in which the Zeeman decelerator is not operated, show that the application of a positive bias voltage, $U_{\mathrm{bias}}$, to the metal pin at the pulsed valve (see Figure \ref{fig:sketchchamber}) can significantly increase the number of metastables produced. Figure \ref{varyp} highlights the change of integrated MCP signals upon application of a 150~V bias voltage as a function of source pressure. Here, signals are shown as ratios with and without 150~V bias voltage to cancel out the pressure dependence at zero bias voltage (cf. Figure \ref{varyp0V}). While there is no change in He and Ne ratios, the signals for N$_2$, Ar and Kr increase very suddenly at certain source pressures and remain, more or less, constant at higher pressures. Figure \ref{varyUbias} shows that the MCP signal intensities for N$_2$, Ar and Kr linearly increase as a function of bias voltage, after a threshold of 15~V for Ar or Kr and 10~V for N$_2$. The gain amounts to more than an order of magnitude at $U_{\mathrm{bias}}$~=~200~V as compared to $U_{\mathrm{bias}}$~=~0~V. In contrast to that, the signal intensities for He and Ne remain unchanged throughout. The observed signal characteristics also correlate with higher electron currents at the pin (Figure \ref{fig:egunchar}) and an increase in light emission during the excitation process (recorded on the MCPs).

Both the threshold behavior for the bias voltage and the step-like increase in signal at a certain source pressure are strong indicators for an electrical breakdown that is typical of a gas discharge, i.e., the generation of an avalanche of charged particles initiated by an electrical current. In principle, there are three main criteria that determine the occurrence of a gas discharge: (1) an initial spark that liberates free electrons which can travel between the cathode and the anode, (2) a sufficient secondary-electron energy to release more electrons from the gas, and (3) a critical gas density (or pressure) 
to allow for sufficient electron-atom/molecule collisions so that more and more secondary electrons and charged particles are created (avalanche process). In our case, condition (1) is fulfilled as soon as electrons from the electron gun reach the valve region. Then, the bias voltage applied to the pin only needs to be sufficiently high so that secondary electrons can be produced, i.e., $U_\mathrm{bias}$ needs to be above the ionization threshold for the gas. As highlighted in Figure \ref{varyUbias}, this characteristic feature is indeed observed for Ar and Kr. In the case of N$_2$, the avalanche process is already initiated at lower bias voltages (inset in Figure \ref{varyUbias}); this may be related to the dissociation of the N--N bond, whose bond-dissociation energy is around 9.8~eV \cite{Darwent1970}.

%The critical pressure for an electrical breakdown (condition (3)) can be estimated from Paschen's law, which describes the relationship between pressure, electrode separation $d_\mathrm{e}$ and applied voltage $U_\mathrm{crit}$ that must be met in order to initiate a gas breakdown, i.e., the sparking of an electrical discharge in a gas \cite{Nasser1971}. These estimates indeed indicate the order of critical pressures that are observed experimentally, with $p$(Kr)~$<$~$p$(Ar)~$<$~$p$(N$_2$). Furthermore, the critical pressure for an electrical breakdown in He and Ne is expected to be higher than the maximum pressure studied in the experiment, providing a sound explanation for the lack of discharge characteristics seen for these gases.

The critical pressure for an electrical breakdown (condition (3)) can be estimated
from Paschen's law, which describes the relationship between pressure, electrode
separation and applied voltage that must be met in order to initiate a gas
breakdown, i.e., the sparking of an electrical discharge in a gas [37]. 
The critical pressures derived from Paschen's law 
follow the order $p$(Kr)~$<$~$p$(Ar)~$\sim$~$p$(N$_2$)~$<$~$p$(Ne)~$<$~$p$(He), which is the same ordering as for the ionization potentials of these gases. The experimental critical pressures for Kr, Ar and N$_2$ in  Figure \ref{varyp} follow the predicted trend, while the lack of a breakdown observed for Ne and He indicates that the critical pressure is beyond
the maximum pressure studied in the
experiment.

% The positively biased metal pin was originally installed to  attract more electrons into the flight path of the supersonic beam. However, the contribution from this effect is small, as seen through simulations in SIMION 8.0 and measurements of the electron beam current at the pin. SIMION simulations also predict that, upon application of a bias voltage, the spatial extent of the electron beam (width of about 2~mm in both the $y$ and the $z$ dimensions) remains unchanged in the excitation region at the valve orifice. However, it comes at the expense of a much broader electron kinetic energy distribution than in the field-free case, from a simulated FWHM of 0.4~eV at U$_\mathrm{bias}$~=~0~V to about 20~eV at U$_\mathrm{bias}$~=~150~V; the total number of electrons is nearly the same at both voltages. However, this broader kinetic energy distribution of the electrons is not reflected in the properties of the supersonic beam, i.e., the measured beam velocities and temperatures are unchanged.

\subsection{\label{sec:boostHe} Discharge excitation in He/Ar and He/Kr mixtures: Pathways to increase the formation of helium in the triplet state}

\begin{figure}[!ht]
\centering
\includegraphics[height=0.206\textheight]{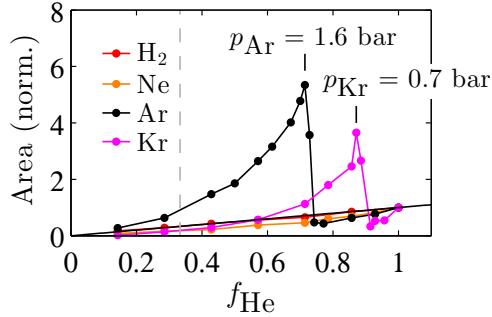}
\caption[Integrated He(2$^3$S$_1$) TOF signal intensity from REMPI detection as a function of He fraction in the initial gas mixture.]{Integrated He(2$^3$S$_1$) TOF signal intensity from (1+1) REMPI detection at 388.97~nm as a function of He fraction, $f_{\mathrm{He}}$, in the initial gas mixture (see legend for carrier gas). The integrated signals for neat He beams ($f_{\mathrm{He}}$~=~1) are used for normalization. $U_{\mathrm{bias}}$~=~150~V in all experiments. The black line indicates a linear decrease in signal intensity expected from a decrease in He concentration. The He/Ar mixing ratio used in Zeeman deceleration experiments is indicated with a dashed gray line. Experimental conditions: $T_0$~=~302~K and $p_0$~=~5.6~bar, valve-to-electron gun trigger delay of 350~$\mu$s.}
\label{HeboostREMPI}
\end{figure}

The discharge mechanism laid out in Section \ref{sec:discharge} was also probed indirectly through measurements of different helium~--~carrier gas mixtures at a bias voltage of 150~V using state-selective REMPI detection of He(2$^3$S$_1$). As can be seen from Figure \ref{HeboostREMPI}, as the fraction of helium is reduced from 1.0, there is a sudden increase in the signal intensity, starting at a particular He/Ar or He/Kr mixing ratio. The measured signal for these gas mixtures is even higher (by up to a factor of 6) than for pure He. A comparison with the results from Section \ref{sec:discharge} suggests that, at pressures $\geq$~1.6~bar for Ar and $\geq$~0.7~bar for Kr, a gas breakdown is initiated which releases an avalanche of secondary electrons.
The avalanche leads in turn  to an increased He(2$^3$S$_1$) signal, which implies that the use of mixed-gas beams may be a beneficial approach for the production of a more intense metastable helium beam for deceleration. The critical pressure for the gas breakdown is lower than for the pure gases, because electronically excited He atoms can further promote the excitation and ionization of the Ar and Kr gas particles. 

The results for mixtures of He with Ne and H$_2$ further support these arguments. No discharge behavior is observed for these gases. Instead, the integrated signal for He/Ne and He/H$_2$ mixtures decreases almost linearly as the He density in the mixture is lowered. This is in good agreement with other experimental results (Figure \ref{varyp0V}) that show a linear dependence of the metastable signal on backing pressure, and this is generally what would be expected if we assume a linear dependence between metastable content and precursor gas density (black line in Figure \ref{HeboostREMPI}). 

Owing to the high energy of the 2$^3$S$_1$ state (19.8~eV \cite{NIST_ASD}), secondary electrons, released in Ar or Kr discharge processes, do not have sufficient energy to excite ground-state He atoms into a metastable level. However, He atoms in the metastable 2$^1$S$_0$ state (20.6~eV \cite{NIST_ASD}) are very efficiently quenched into the lower-lying, metastable triplet state through collisions with thermal electrons \cite{Phelps1955}. As the cross section for the singlet-to-triplet conversion process is very high (3$\times$10$^{-14}$cm$^{2}$ \cite{Phelps1955}), this is probably the dominant process leading to the observed boost in He(2$^3$S$_1$) signal intensity. The results are also consistent with the generally high singlet-to-triplet He content observed for electron-impact sources, e.g., 2$^1$S$_0$:2$^3$S$_1$~=~7:1 for the source from Brutschy and Haberland \cite{Brutschy1977}. In fact, this conversion process has been used in the past to ensure large triplet-to-singlet ratios in metastable He beams \cite{Schmeltekopf1970}. To confirm this mechanism, validation measurements could be carried out in which the signal intensity of He($^1$S$_0$) is determined, e.g., through a (1+1) REMPI via the 3$^1$P$_1$ state at 501.71~nm \cite{Haberland1987}.

The results in Figure \ref{HeboostREMPI} also illustrate that the He(2$^3$S$_1$) signal decreases non-linearly as the Ar or Kr partial pressure in the mixture is increased. This process is not yet fully understood. It may be related to the de-excitation of He(2$^3$S$_1$) owing to the Penning ionization of Ar/Kr \cite{Siska1993} or it could be related to an increased transverse spreading of the beams due to lower beam velocities at higher carrier gas pressures.

For He(2$^3$S$_1$) Zeeman deceleration experiments, a 1:3 He/Ar mixing ratio was used to attain an initial beam velocity of around 500~m/s. With the nozzle at room temperature, the signal enhancement from the discharge mechanism for these experimental conditions is only marginal (dashed gray line in Figure \ref{HeboostREMPI}). However, the signal gain can be easily increased by choosing a different overall source pressure. Owing to the change in discharge characteristics, the signal gain will also depend on the chosen source temperature.
To find optimum working conditions, the source pressure would have to be varied systematically at the chosen valve temperature.

\section{\label{sec:concl}Conclusions}

The experiments presented here demonstrate the production of a beam of Helium atoms in the 2$^3$S$_1$ state for Zeeman deceleration, and  show that the velocity of He atoms in this state can  be considerably reduced with a short 12-stage Zeeman decelerator - up to a 40\% reduction in kinetic energy. As far as we are aware, this is the first successful Zeeman deceleration experiment of metastable atomic helium. The experimental data also show that the pulse duration for electron impact should be varied according to the phase-space acceptance of the Zeeman decelerator in order to obtain decelerated beams with well-defined final velocity distributions.

Our results indicate that a discharge-type process, induced by applying a positive voltage to a metal pin at the pulsed valve, can lead to  increased formation of excited-state species as observed for Ar, Kr and N$_2$ gases at room temperature. Singlet-to-triplet conversion by secondary electrons from a discharge in Ar or Kr can be used to significantly increase the number of He atoms in the metastable 2$^3$S$_1$ state. As the metal pin used in the present setup was originally not designed with the purpose of creating a discharge, it remains to be seen whether metastable production can be further increased by the use of a different anode material or a more homogeneous electric field, e.g., via a metal plate parallel to the electron gun.

The injection seeding of a discharge with electrons -- the opposite approach to what is described here -- has already been successfully used for the stable operation of an intense, pulsed metastable He source \cite{Halfmann2000}. Nevertheless, it is not clear whether low-velocity supersonic beams of He(2$^3$S$_1$) could be achieved using such a conventional discharge-type setup, as seeding in a heavier carrier gas may lead to an efficient quenching of the metastable helium signal \cite{Raunhardt2009}.

Our findings provide a more detailed understanding of the mechanisms that govern metastable formation in a combined electron-gun-discharge source, and may eventually lead to the development of even more intense sources of metastable atoms and molecules.
This work also shows that a pulsed electron gun is a suitable source of metastable atoms and molecules for Zeeman deceleration. This source has potential value for applications in collision experiments, for example, in Penning ionization studies at high-energy resolution, and possibly in trapping and laser cooling schemes.

\ack

We are grateful for technical support from James N Bull (now at Durham University, UK), Howard Lambourne and Neville Baker (PTCL workshops, Oxford, UK). K.D. acknowledges financial support from the Chemical Industry Fund (FCI, Germany) through a Kekul\'{e} Mobility Fellowship, and through a grant from the Simms Foundation (Merton College, Oxford, UK). This work is financed by the Engineering and Physical Sciences Research Council (U.K.) EPSRC(GB) under Project Nos. EP/G00224X/1 and EP/1029109.

\bibliography{Zeeman3bib}

\end{document}